# A VALIDATED COUPLED THREE-DIMENSIONAL HYDRODYNAMIC AND SPECTRAL WIND-WAVE MODEL FOR THE WESTERN NORTH ATLANTIC OCEAN

Maria Venolia[1], Reza Marsooli[1]*, Jaime R. Calzada[2].

1. Stevens Institute of Technology, Hoboken, NJ 08540

2. Virginia Institute of Marine Science, Gloucester Point, VA

**Abstract**

Wind-wave and ocean current interactions affect critical coastal and oceanic processes, yet modeling these interactions presents significant challenges. The western North Atlantic Ocean provides an ideal test environment for coupled hydrodynamics and wind wave models, thanks to its energetic surface currents such as the Gulf Stream. This study evaluates a high-resolution coupled SCHISM+WWM III model, utilizing NOAA's "STOFS-3D-Atlantic" computational mesh, while incorporating three-dimensional baroclinic dynamics to account for density stratification effects. We evaluate the model's calculated water level and tidal predictions against NOAA tide gauge measurements during December 2016. The coupled model demonstrates robust skills in reproducing tidal constituents, non-tidal components, and total water level predictions along the U.S. East and Gulf of Mexico Coasts. In addition, we systematically evaluate three wave physics parameterizations (Ardhuin, Makin and Stam, and Cycle Three) in the spectral wave model to quantify their effects on the modeled wave characteristics. This validated modeling framework enhances our ability to understand and predict complex coastal and oceanic processes, offering significant applications for coastal management, maritime operations, and climate adaptation planning throughout the western North Atlantic region.

1. **Introduction**

Oceanic and coastal environments are governed by the complex interactions of surface gravity waves and currents, which have been extensively studied (ASM

Alauddin and Marsooli, 2024; Echevarria et al., 2019; Jonsson et al., 1970; Markina et al., 2018; Marsooli et al., 2017; Roland et al., 2012). Wave-current interactions occur through several key physical mechanisms, including the transfer of wave momentum to the mean flow, wave-enhanced surface roughness and turbulence, as well as modified bottom stress in shallow waters due to the wave orbital motion (Grant and Madsen, 1979; Janssen, 1991; Longuet-Higgins and Stewart, 1962). Additional processes, such as Stokes drift and wave refraction by currents, further contribute to wave-current coupling (Ardhuin et al., 2017, 2009; Li and Chabchoub, 2024). These processes are further complicated in baroclinic systems, where stratification due to temperature and salinity gradients introduces additional variability in current dynamics and consequently wave-current interactions (Chen et al., 2023; Li et al., 2024). Such interactions can affect driving forces of a wide range of coastal and oceanic processes and activities, including coastal erosion and shoreline evolution (Armstrong and Flick, 1989; Franco-Ochoa et al., 2020), renewable wave energy harvesting (Mi et al., 2022; Scruggs and Jacob, 2009), operations in the oil and gas industry (Bijker and Leeuwestein, 1984; Tuhaijan et al., 2011), and maritime navigation (Echevarria et al., 2021).

The western North Atlantic Ocean presents a particularly demanding environment for accounting for wave-current interactions, due to its complex bathymetry, energetic western boundary currents, e.g., the Gulf Stream, and highly variable wave climate (Lin and Sheng, 2023). This region experiences significant seasonal and interannual variability in wave conditions (Jamous and Marsooli, 2023; Venolia et al., 2024; Venolia and Marsooli, 2025) and surface current patterns (Andres, 2021; Bromirski and Cayan, 2015). Previous modeling efforts in this region have highlighted the importance of accurately representing both coastal and open ocean processes to capture the full spectrum of wave-current dynamics (Cui et al., 2024).

The prediction of water levels, waves, and coastal flooding in the western North Atlantic has often treated waves and currents (hydrodynamics) separately, overlooking their interactions. For instance, NOAA's Global Surge and Tide Operational Forecast System (STOFS-2D-Global) uses the stand-alone ADCIRC model for hydrodynamic forecasts, while NOAA's wave forecasts are generated by stand-alone spectral models such as SWAN in the Nearshore Wave Prediction System (NWPS) and WAVEWATCH III in the Global Ensemble Forecast System - Wave (GEFS-Wave). Recognizing the significance of wave-current interactions on water levels and wave conditions, there has been a growing adoption of two-way coupled hydrodynamic and wave models for coastal flood forecasting. However, this coupling has predominantly involved two-dimensional hydrodynamic models (e.g., Al Azad and Marsooli, 2024). One example is the Coastal Emergency Risks Assessment (CERA) system, which couples the two-dimensional ADCIRC hydrodynamic model with the SWAN wave model. Moreover, most wave-coupling efforts in the western North Atlantic basin have been based on barotropic hydrodynamic modeling, whereas three-dimensional (3-D) baroclinic dynamics to capture stratification, buoyancy-driven currents, and vertical structure are neglected.

The Semi-implicit Cross-scale Hydroscience Integrated System Model (SCHISM) is a 3-D coastal and ocean circulation model that has demonstrated robust cross-scale hydrodynamic modeling capabilities (Zhang et al., 2016). It has been extensively utilized for various applications, including flooding during hurricanes (Ye et al., 2019; Zhang et al., 2020) and baroclinic ocean dynamics (Ye et al., 2021). SCHISM serves as the hydrodynamic core for NOAA 3-D Surge and Tide Operational Forecast System for the Atlantic Basin (STOFS-3D-Atlantic), which provides water level forecasts for the U.S. East Coast and the Gulf of Mexico. While Cui et al., (2024) have comprehensively validated STOFS-3D-Atlantic's mesh capability to predict total water levels with high accuracy, their model doesn't incorporate waves and their interactions with currents.

Despite significant advances in coupled wave-current modeling, a comprehensive review of the literature reveals a notable gap in the field: the absence of a large-scale, cross-scale, three-dimensional fully coupled baroclinic wave model for the western North Atlantic region. While there exist examples of three-dimensional baroclinic models (e.g., ESTOFS) and two-dimensional large-scale coupled wave models (Abdolali et al., 2020; Roland et al., 2012), a modeling framework that integrates both these important components has not yet been demonstrated. Our approach leverages SCHISM's robust capabilities for cross-scale three-dimensional modeling to develop what appears to be the first fully coupled spectral wave model at this scale and resolution that incorporates three-dimensional baroclinic dynamics. This advancement represents a significant step forward in capturing the complex wave-current interactions that govern coastal and oceanic processes in this region.

This study aims to address the above-mentioned gap by evaluating the performance of a high-resolution coupled SCHISM and Wind-Wave Model III (WWM III) in the western North Atlantic Ocean based on NOAA's STOFS-3D-Atlantic computational mesh. We evaluate the model accuracy in simulating water levels and waves, incorporating full baroclinic dynamics to account for density variations due to temperature and salinity gradients. The performance of the model is evaluated for different wave physics parameterization methods that are available in the wave model. Model evaluation is performed using in-situ measurements from National Oceanic and Atmospheric Administration (NOAA) tide gauges and National Data Buoy Center (NDBC) buoys, comparing observed water levels, significant wave heights, mean wave periods, and directions during December 2016.

2. **Methodology**
2.1 **Model Description**
2.1.1 **SCHISM**

We conduct baroclinic, three-dimensional hydrodynamic simulations using SCHSIM, which is an open-source, unstructured-grid, hydrodynamic modeling system for coastal and ocean environments across various spatial and temporal scales (Zhang et al., 2016). SCHISM is a derivative product of SELFE (Semi-implicit Eulerian-Lagrangian Finite-Element) (Zhang and Baptista, 2008) with many enhancements. The governing equations solved by the model include:

Momentum equation:

$$\frac{Du}{Dt} = \frac{\partial}{\partial z}\left(v\frac{\partial u}{\partial z}\right) - g\nabla\eta + F \quad (1)$$

Continuity equation in 3D form:

$$\nabla u + \frac{\partial w}{\partial z} = 0 \tag{2}$$

Transport equation:

$$\frac{\partial C}{\partial t} + \nabla(uC) = \frac{\partial}{\partial z}\left(\kappa \frac{\partial C}{\partial z}\right) + F_h \tag{3}$$

where $\nabla = (\partial\,\partial x, \partial\,\partial y)$ is the horizontal gradient operator, D/Dt is the material derivative, (x,y) are the horizontal Cartesian coordinates, z is the vertical coordinate, t is time, η(x, y, t) is the free-surface elevation, u (x, y, z, t) is the horizontal velocity, with Cartesian components (u,v), w is the vertical velocity, F is other forcing terms in the momentum equation including horizontal viscosity terms and wave-induced forcing described in section 2.1.3, g is the acceleration of gravity, C is a tracer concentration (e.g., salinity, temperature, sediment etc.), ν is the vertical eddy viscosity, κ is the vertical eddy diffusivity for tracers, and $F_h$ is the horizontal diffusion and mass sources/sinks (Zhang et al., 2016).

Equations (1) through (3) form a differential system which is closed by applying one of the turbulence closure models available within SCHISM (Umlauf and Burchard, 2003), like k-ε (Mohammadi and Pironneau, 1993), k-ω (Umlauf et al., 2003) and Mellor-Yamada (Ezer, 2000). In this study, we utilize the k-ε turbulence model (with Kantha & Clayson stability function).

### 2.1.2 WWM III

The governing equation of the third-generation spectral wave model WWM III is the wave action Equation (Roland et al., 2012) as follows

$$\frac{\partial}{\partial t}N + \nabla_x(\dot{X}N) + \frac{\partial}{\partial \sigma}(\dot{\theta}N) + \frac{\partial}{\partial \dot{\theta}}(\dot{\sigma}N) = S_{tot} \tag{4}$$

with the wave action is defined as

$$N_{(t,X,\sigma,\theta)} = \frac{E_{(t,X,\sigma,\theta)}}{\sigma} \tag{5}$$

where E being the variance density of the sea level elevations, σ the relative wave frequency, and θ the wave direction. The advection velocities $\dot{X}$, $\dot{\theta}$ and $\dot{\sigma}$ in the
5

different phase spaces and the source function $S_{tot}$ are all influenced by currents and water depths from SCHISM (Schloen et al., 2017). The source term $S_{tot}$ is defined below, accounting for wind energy input, wave-wave interactions, and energy dissipation due to whitecapping, depth-induced breaking, and bottom friction, which can be calculated using different wave physics packages described in section 2.1.3.

$$S_{tot} = \underbrace{S_{in}}_{wind\ energy\ input} + \underbrace{S_{nl4} + S_{nl3}}_{nonliner\ interaction} + \underbrace{S_{ds}}_{whitecapping} + \underbrace{S_{br}}_{wave\ breaking} \quad (6)$$
$$+ \underbrace{S_{bf}}_{bottom\ friction}$$

### 2.1.3 Coupled SCHISM + WWM III model

SCHISM compilation allows users to activate the WWM III module, since WWM III is recast as a subroutine within SCHISM at the source code level. The two models share the same domain decomposition in the parallel MPI implementation (Schloen et al., 2017). According to Roland et al., (2012) the coupled model has proven to be both accurate and robust. During the information exchange, the water surface elevation and velocity are passed from SCHISM to WWM III (Schloen et al., 2017). WWM III provides SCHISM with wave radiation stresses as well as friction (shear) velocities, which are used to specify the surface stress in the circulation model. This coupling approach has been successfully applied across various coastal environments, from storm surge applications (Liu et al., 2012) to complex inner sea systems (Cavaleri et al., 2014), demonstrating the model's capability to capture wave-current interactions under diverse conditions.

In SCHISM, the wave-induced radiation stresses proposed by Longuet-Higgins and Stewart, (1964) are incorporated in the F forcing term of equation (1) (Schloen et al., 2017). This F term according to (Roland et al., 2012) is

$$F = \nabla \cdot (\mu \nabla \vec{u}) - f\vec{k_z} \times \vec{u} - \frac{1}{\rho_0} \nabla p_A + ag \nabla \varphi + \vec{R_s} \quad (7)$$

where $\vec{k_z}$ is a unit vector of the z-axis, $f$ is the Coriolis factor, $a$ is the effective earth elasticity factor, φ is the earth tidal potential, μ is the horizontal eddy viscosity and $p_A$ is the atmospheric pressure. $\overrightarrow{R_s}$ is the radiation stress term, which according to Longuet-Higgins and Stewart (1964) is

$$\overrightarrow{R_s} = (R_{sx}, R_{sy}) \tag{8}$$

$$R_{sx} = -\frac{1}{\rho_0 H}\frac{\partial S_{xx}}{\partial x} - \frac{1}{\rho_0 H}\frac{\partial S_{xy}}{\partial y} \tag{9}$$

$$R_{sy} = -\frac{1}{\rho_0 H}\frac{\partial S_{yy}}{\partial y} - \frac{1}{\rho_0 H}\frac{\partial S_{xy}}{\partial x} \tag{10}$$

The components of the radiation stress tensor are defined according to Battjes (1972) as

$$S_{xy} = \int_0^{2p}\int_0^{\infty} N(\sigma,\Theta) \cdot \sigma \cdot \frac{c_g(\sigma)}{c_p(\sigma)} \sin(\Theta)\cos(\Theta)\, d\Theta\, d\sigma \tag{11}$$

$$S_{xx} = \int_0^{2p}\int_0^{\infty} N(\sigma,\Theta) \cdot \sigma \cdot \left[\frac{c_g(\sigma)}{c_p(\sigma)}(cos^2(\Theta)+1) - \frac{1}{2}\right] d\Theta\, d\sigma \tag{12}$$

$$S_{yy} = \int_0^{2p}\int_0^{\infty} N(\sigma,\Theta) \cdot \sigma \cdot \left[\frac{c_g(\sigma)}{c_p(\sigma)}(sin^2(\Theta)+1) - \frac{1}{2}\right] d\Theta\, d\sigma \tag{13}$$

In the above equations $H = h + \eta$, with h being bathymetric depth, which is the total water depth obtained from SCHISM, and $c_g$ and $c_p$ are the group and phase velocities respectively (Schloen et al., 2017). Wave effects in SCHISM are also accounted for through friction velocities, which are used to calculate the water surface stress based on the wave-depended surface roughness and according to Pond and Pickard, (1983).

In WWM III, the advection velocities presented in equation (4) account for the presence of currents as the following (Schloen et al., 2017) :

$$\dot{X} = \frac{d\vec{X}}{dt} = \frac{d\omega}{dk} = \vec{c}_g + \vec{U}_{A(k)} \tag{15}$$

$$\dot{\Theta} = \frac{1}{k}\frac{\partial \sigma}{\partial H}\frac{\partial H}{\partial m} + \vec{k}\frac{\partial \vec{U}_{A(k)}}{\partial s} \tag{16}$$

$$\dot{\sigma} = \frac{\partial \sigma}{\partial H}\left(\frac{\partial H}{\partial t} + \vec{U}_A \cdot \nabla_{\vec{x}} H\right) - \vec{c}_g \vec{k}\frac{\partial \vec{U}_{A(k)}}{\partial s} \tag{17}$$

where s is the coordinate along the wave propagation direction, and m is perpendicular to it. $\nabla_{\vec{x}}$ is the gradient operator and $\vec{U}_A$ is the surface current obtained from SCHISM.

Equation (15) represents the propagation in space and accounts for shoaling including the effects of currents, while equation (16) represents the depth- and current-induced refraction, and equation (17) accounts for the shifts in frequency due to variations in depth and currents (Schloen et al., 2017).

### 2.1.4 Wave physics parameterizations in WWM III

In our evaluation of wave modeling performance of WWM III, we examine three distinct wave physics parameterizations: "Cycle Three", "Makin and Stam", and "Ardhuin". We compare model outputs against observational data from NOAA NDBC stations distributed throughout our study domain. To quantify model performance, we calculated bias and root mean square error (RMSE) between modeled results and NDBC data for significant wave height, mean wave period, and wave direction. This comparative assessment reveals differences in how each parameterization scheme represents complex wave processes.

The "Cycle Three" parameterization represented a significant advancement in wave modeling when introduced by the WAMDI Group, (1988). Unlike previous generation (first and second generation) models that imposed restrictive spectral shapes, Cycle Three was the first method to integrate the basic transport equation without additional assumptions about spectral form. It explicitly prescribed three source functions: wind input based on Snyder et al., (1981) and Komen et al., (1984), nonlinear transfer through a discrete interaction approximation, and a white-capping dissipation source function. This third-generation model successfully addressed previous limitations in simulating complex wind-seas and the transition between wind-sea and swell conditions. Cycle Three's verification across numerous hindcast studies, including North Atlantic storms and Gulf of Mexico hurricanes (WAMDI Group, 1988), demonstrated its robust performance in predicting fetch-limited wave growth and developing Pierson-Moskowitz spectra under various wind conditions.

The "Makin and Stam", (2003) formulation represents an alternative approach to wave physics parameterization based on advanced wind-over-waves coupling theory. Their model addresses air-sea interactions through a resistance law that explicitly accounts for the interplay between viscous stress and wave-induced stress components. A distinctive feature of this approach is its physics-based representation of how wave age influences drag coefficients, with younger, steeper seas generating greater sea drag than fully developed wave fields (Makin and Kudryavtsev, 2002; Makin and Stam, 2003). When implemented in wave forecasting systems, this formulation has been shown to reduce biases in wave height predictions compared to traditional approaches, particularly in higher wind speed conditions (Zweers et al., 2012).

Ardhuin et al., (2010) developed a physically consistent framework without predetermined spectral constraints. Their approach integrated nonlinear swell dissipation proportional to wave steepness, implemented threshold-based saturation criteria for breaking, and introduced a cumulative term that accounts for the dissipation of shorter waves by longer breaking waves.

## 2.2 Study Area and Computational Domain

The present study employs the computational mesh from the NOAA 3-D Surge and Tide Operational Forecast System for the western North Atlantic basin (STOFS-3D-Atlantic). This system, both operational and experimental, is utilized for storm surge and tide forecasting and is undergoing active development and testing by the National Ocean Service. STOFS-3D-Atlantic employs SCHISM as its hydrodynamic model core (https://registry.opendata.aws/noaa-nos-stofs3d, last accessed May 2024). This computational domain encompasses the U.S. East Coast and the Gulf of Mexico, reaching as far north as the Gulf of Maine (Figure 1). To ensure compatibility with WWM III, we converted quadrilateral elements within the STOFS-3D-Atlantic mesh into triangles using a Fortran-based triangulation script included in the SCHISM utility codes. The modified unstructured grid consists of 2,926,236 nodes and 5,654,157 elements. The mesh resolution ranges from about 2 to 7 km at the open deep ocean to 1.5 to 2 km near the coastlines, and approximately 600m within floodplain areas (Experimental STOFS Web Portal). Although the present study focuses on waves and water levels in oceanic and coastal waters, we chose to retain floodplain areas for future coastal flood modeling applications.

The vertical grid configuration employed in the present research incorporates 20 sigma S layers, structured to handle the vertical discretization of the water column effectively. Unlike typical layered models that may use a combination of Z and S layers, this configuration employs a pure S layer approach, adapting it specifically for uniformity and continuity across the water column. While LSC2 (Localized Sigma-Square) grids exist and offer significant advantages in computational speed, the choice of S grid configuration for this study stems from the use of a retriangulated version of STOFS-3D-Atlantic mesh. Although S grids generally offer lower computational performance compared to LSC2 grids, they provide the advantage of being relatively straightforward to generate while still maintaining adequate representation of the water column. The S layer configuration begins at a depth of -8,428 meters, marking the deepest point of the modeling domain. By employing only sigma S layers, the model ensures that vertical variations in the water column are captured with high fidelity, facilitating precise simulations of bottom-controlled processes (Huang et al., 2022).

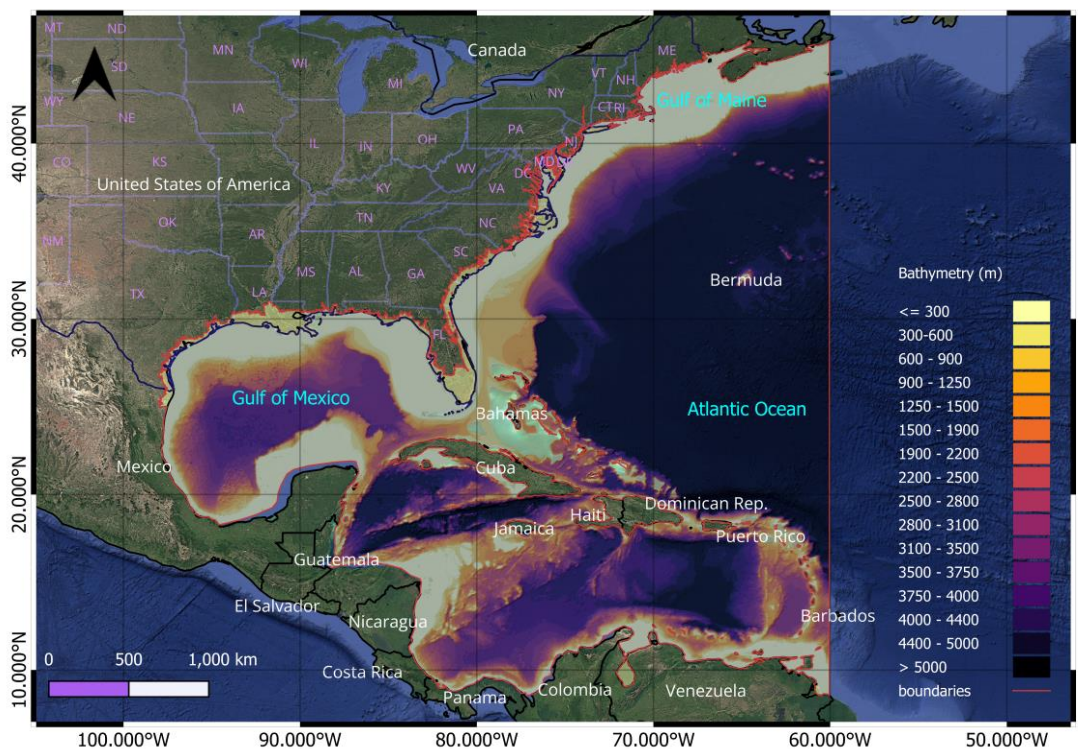

*Figure 1. NOAA STOFS-3D-Atlantic computational domain.*

### 2.3 Model configuration

A two-way coupling approach is adopted, where information exchange between SCHISM and WWM III occurs at a pre-specified 20 minute interval (equivalent to 8 model time steps). The simulation period covers December 2016, with a time step of 150 seconds. This simulation period was selected due to its variability in wave and swell conditions along the U.S. East and Gulf Coasts (ASM Alauddin and Marsooli, 2024). Model initial conditions for surface elevation and water column velocity, temperature and salinity are obtained from the HYCOM ocean modeling system (https://www.hycom.org/ last accessed December 2024) which provides global oceanic conditions at a spatial resolution of 0.08 degrees (approximately 8 km). A two-week warm-up period (during November 2016) was implemented, to allow development of the circulation patterns, prior to the December 2016 analysis period, ensuring more reliable comparison with observational data.

Bottom friction is implemented through spatially variable bottom drag coefficients. This coefficient was determined through multiple calibration runs (testing values of 0.015, 0.025, and 0.035 across the entire domain - excluding the floodplains) and evaluating performance against tidal gauge measurements of total water level. The final spatial distribution was manually optimized based on regional hydrodynamic response: a drag coefficient of 0.0035 in the Gulf of Maine and its continental shelves (where the tidal amplitude is large), 0.0015 in the Gulf of Mexico, 0.0025 in coastal and deep ocean regions off the East Coast, and 0.025 across floodplain zones.

#### 2.3.1　Oceanic and atmospheric forcings

The model is forced using winds and atmospheric pressure at its sea surface boundary and using tides, currents, temperature and salinity fluxes, and directional wave spectra at its open ocean boundary. Tidal elevation and velocity conditions are imposed at computational nodes along the open ocean boundary of SCHISM, using 8 major tidal constituents (M2, S2, N2, K2, K1, O1, P1, and Q1). The tidal boundary conditions were generated based on the 2014 version of the Finite Element Solution tide dataset (2014 FES) (Lyard et al., 2021), using the "pyschism" tool (Calzada et al., 2023) .

The 2014 FES atlas is the latest release of FES global ocean tidal atlases aiming to improve tidal predictions. Compared to earlier releases, the tidal constituent's spectrum is significantly extended while at the same time the overall resolution is augmented.

Atmospheric conditions, including pressure and wind fields, are based on hourly re-analysis data acquired from ECMWF (European Centre for Medium-Range Weather Forecasts) ERA5 dataset (Hersbach et al., 2017), which have a horizontal resolution of approximately 31 km globally. Boundary conditions for elevation, velocity, salinity, and temperature are obtained from HYCOM.

Two-dimensional direction-frequency spectral wave data obtained from ECMWF ERA5 global reanalysis is prescribed at the open boundaries of WWM III. These spectral boundary conditions allow for accurately representing incoming wave energy, particularly swell waves that originate from remote meteorological systems outside our computational domain.

Although SCHISM has the capability to incorporate river inputs as sources and sinks from various discharge data sources (such as the NOAA National Water Model (Cosgrove et al., 2024) for U.S. watersheds or other regional/global databases for international rivers), riverine discharges were not included in the present study. The focus of this investigation is on open-ocean processes driven by tides, waves, and atmospheric forcings, rather than estuarine dynamics. Consequently, when we use the term "total water level" in this paper, we refer specifically to the combined effect of astronomical tides, meteorological forcing (winds and pressure), and wave-induced setup, without the contribution of riverine discharges.

3. Results and Discussion

In this section, we evaluate model performance by comparing outputs against observational data. Model validation is conducted using in-situ measurements from NOAA tide gauges (Table 1) and NDBC buoys (Table 2).

*Table 1. List of NOAA tide gauges ordered form North to South along the U.S. East coast and from east to West within the Gulf of Mexico.*

| Station ID | Longitude | Latitude | Name |
|---|---|---|---|
| 8411060 | 67.20466 W | 44.65702 N | Cutler Farris Wharf, ME |
| 8413320 | 68.20427 W | 44.39219 N | Bar Harbor, ME |
| 8418150 | 70.24417 W | 43.65806 N | Portland, ME |
| 8423898 | 70.71055 W | 43.07138 N | Fort Point, NH/ME |
| 8443970 | 71.05028 W | 42.35390 N | Boston, MA |
| 8447930 | 70.67111 W | 41.52361 N | Woods Hole, MA |
| 8452660 | 71.32614 W | 41.50433 N | Newport, RI |
| 8454049 | 71.41000 W | 41.58694 N | Quonset Point, RI |
| 8465705 | 72.90833 W | 41.28333 N | New Haven, CT |
| 8467150 | 73.18396 W | 41.17581 N | Bridgeport, CT |
| 8510560 | 71.95944 W | 41.04833 N | Montauk, NY |
| 8516945 | 73.76571 W | 40.81134 N | Kings Point, NY |
| 8518750 | 74.01417 W | 40.70055 N | The Battery, NY |
| 8531680 | 74.00938 W | 40.46683 N | Sandy Hook, NJ |
| 8534720 | 74.41805 W | 39.35666 N | Atlantic City, NJ |
| 8537121 | 75.37668 W | 39.30538 N | Ship John Shoal, NJ/DE |
| 8557380 | 75.11927 W | 38.78283 N | Lewes, DE |
| 8573364 | 76.24456 W | 39.21344 N | Tolchester Beach, MD |
| 8574680 | 76.57944 W | 39.26694 N | Baltimore, MD |
| 8575512 | 76.41536 W | 38.95742 N | Annapolis, MD |
| 8577330 | 76.37790 W | 38.33005 N | Solomons Island, MD |
| 8632200 | 75.98844 W | 37.16519 N | Kiptopeke, VA |
| 8636580 | 76.29665 W | 37.60103 N | Windmill Point, VA |
| 8637689 | 76.25144 W | 37.13055 N | Yorktown USCG Training |
| 8638610 | 76.33759 W | 36.94359 N | Sewells Point, VA |
| 8638863 | 76.11790 W | 36.96931 N | Chesapeake Bay Bridge Tunnel, |
| 8651370 | 75.74669 W | 36.1833 N | Duck, NC |
| 8658163 | 34.21333 W | 34.21333 N | Wrightsville Beach, NC |
| 8721604 | 80.59305 W | 28.41583 N | Trident Pier, Port Canaveral, FL |
| 8722670 | 80.03416 W | 26.61277 N | Lake Worth Pier, FL |
| 8724580 | 81.81002 W | 24.56980 N | Key West, FL |
| 8726724 | 82.83166 W | 27.97833 N | Clearwater Beach, FL |
| 8725110 | 81.82305 W | 26.12877 N | Naples, Gulf of Mexico, FL |
| 8729210 | 85.66444 W | 30.14972 N | Panama City Beach, FL |
| 8735180 | 88.07500 W | 30.25000 N | Dauphin Island, AL |
| 8761305 | 89.67321 W | 29.88040 N | Shell Beach, LA |
| 8768094 | 93.33654 W | 29.75014 N | Calcasieu Pass, LA |
| 8772447 | 95.30250 W | 28.94330 N | Freeport, TX |
| 8775870 | 97.21670 W | 27.58000 N | Bob Hall Pier, Corpus Christi, |

*Table 2. List of NOAA NDBC buoy stations from North to South along the U.S. East coast and from West to East within the Gulf of Mexico.*

| *Buoy Station* | Longitude | Latitude | Depth (m) |
|---|---|---|---|
| 44027 | 67.300 W | 44.283 N | 191.4 |
| 44005 | 69.127 W | 43.201 N | 176.8 |
| 44017 | 72.049 W | 40.693 N | 48 |
| 44025 | 73.164 W | 40.251 N | 36.3 |
| 44066 | 72.644 W | 39.618 N | 77 |
| 41002 | 74.936 W | 31.759 N | 3784 |
| 41048 | 69.573 W | 31.831 N | 5394 |
| 41049 | 63.012 W | 27.545 N | 5413 |
| 42019 | 95.345 W | 27.910 N | 83.5 m |
| 42002 | 93.646 W | 26.055 N | 3088 m |
| 42001 | 89.662 W | 25.926 N | 3200 m |
| 42003 | 85.616 W | 25.925 N | 3273 m |

### 3.1 Astronomical Tidal Harmonic Analysis

Tide-induced cyclical patterns in sea levels can be mathematically decomposed into constituent components - otherwise known as harmonic constituents. Through harmonic analysis of water level data (typically requiring at least 30 days of measurements (Parker, 2007), these constituent components can be isolated and quantified. In order to extract the harmonic constituents from the model time series data, we use the T_TIDE software package (Pawlowicz et al., 2002), which implements classical harmonic analysis methods to decompose water level signals into their constituent tidal components. The package performs least-squares fitting of sinusoids at specific tidal frequencies to the data, providing amplitude and phase information for each constituent along with statistical confidence intervals. This approach allows for the separation of deterministic tidal signals from non-tidal components in the water level records.

Figure 2 shows the amplitude, phase, and complex error of M2 constituent for the tidal gauge locations along the East and Gulf coasts listed in Table 1, while Figure 3 shows similar results but for K1 constituent. The complex error is defined as the Root-Mean-Square Error (RMSE) that includes both amplitude and phase error (Cui et al., 2024; Huang et al., 2022):

$$RMSE = (0.5 * ((A_0 cos P_0 - A_m cos P_m)^2 \qquad (18)$$
$$+ (A_0 sin P_0 - A_m sin P_m)^2))^{1/2}$$

where $A$ is amplitude, $P$ is phase, and the subscripts $o$ and $m$ refer to observation and model, respectively.

Most of the stations along the U.S. East coast are dominated by the principal lunar semi-diurnal M2 constituent, while lunar diurnal K1 is the dominant constituent for many stations along the Gulf of Mexico (Cui et al., 2024). In most stations, we notice strong agreement between the model and observations, both in terms of amplitude and phase. The mean complex error across all stations for M2 is 0.074 m, while for K1 it is 0.027 m.

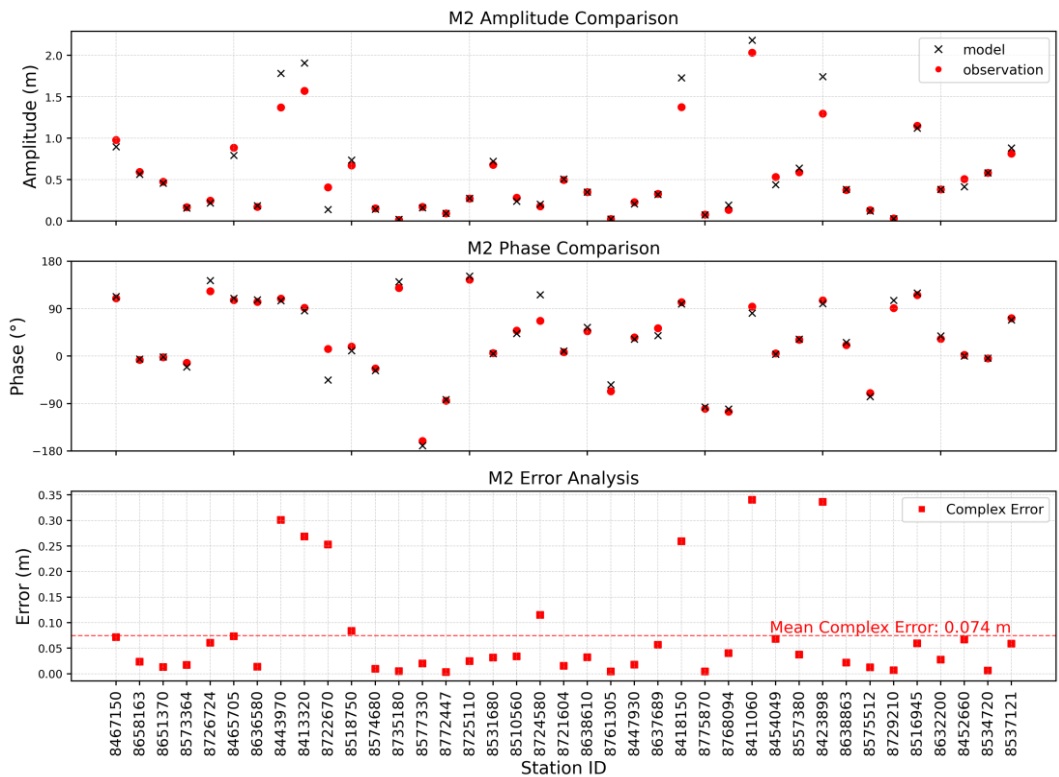

*Figure 2. Comparison between modeled and observed M2 tidal constituent parameters, showing amplitude, phase, and complex error analysis.*

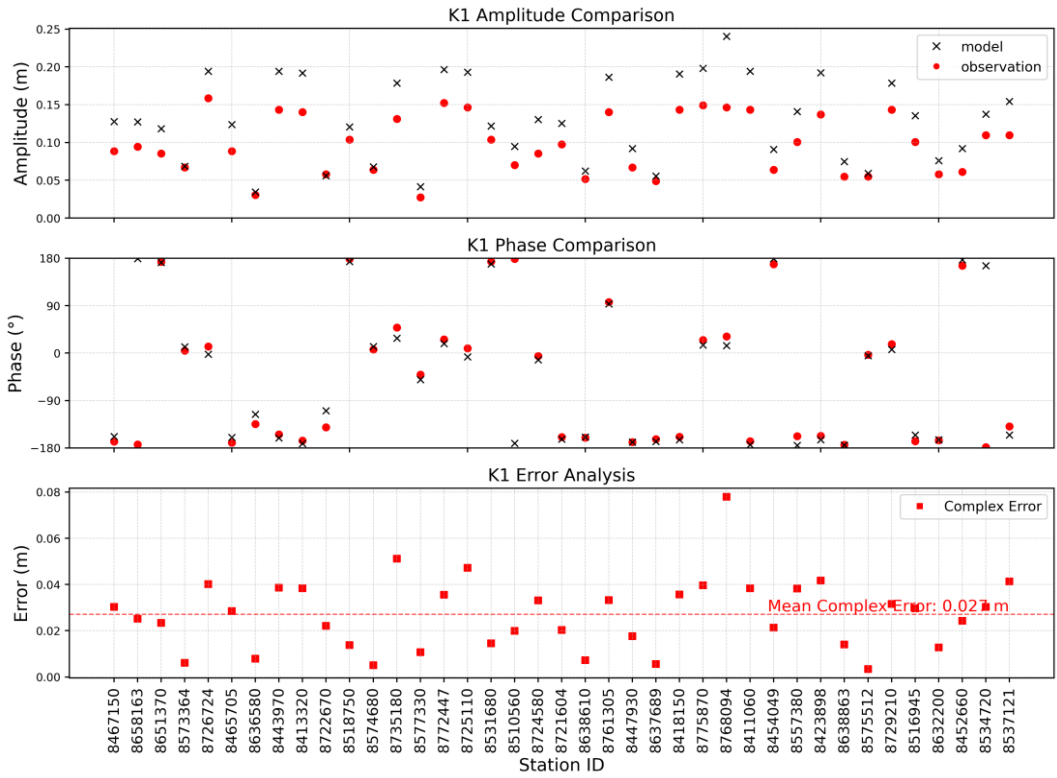

*Figure 3. Comparison between modeled and observed K1 tidal constituent parameters, showing amplitude, phase, and complex error analysis.*

### 3.2 Non-tidal elevation

We followed (Cui et al., 2024) to calculate the non-tidal elevation using a low-pass filter with a two-day cutoff. Low-pass filtering allows us to remove noise from the total water level, which consists of both tidal and non-frequency (non-tidal) oscillations. In the case of a two-day cutoff, all oscillations with periods shorter than two days are eliminated, while preserving longer-term variations caused by weather systems and other non-tidal phenomena. This approach enables the separation of deterministic tidal signals from stochastic meteorological and oceanographic influences. Figures 4 and 5 illustrate the spatial distribution of the non-tidal Root Mean Square Error (RMSE) and bias, respectively, for the simulation period of December 2016. The minimum non-tidal RMSE is 0.07 m while the maximum is 0.30 m (Table 3). However, the majority of the stations exhibit non-tidal RMSE ranging between 0.10 and 0.20 m, while the mean across all stations is approximately 0.14 m. At the same time, the model is generally negatively biased during the simulation period, with the bias at most stations ranging between -0.10 m and -0.15 m, with a mean of approximately -0.12 m. The minimum nontidal bias in our domain is as low as -0.04 m while the maximum reaches around -0.29 m (Table 3).

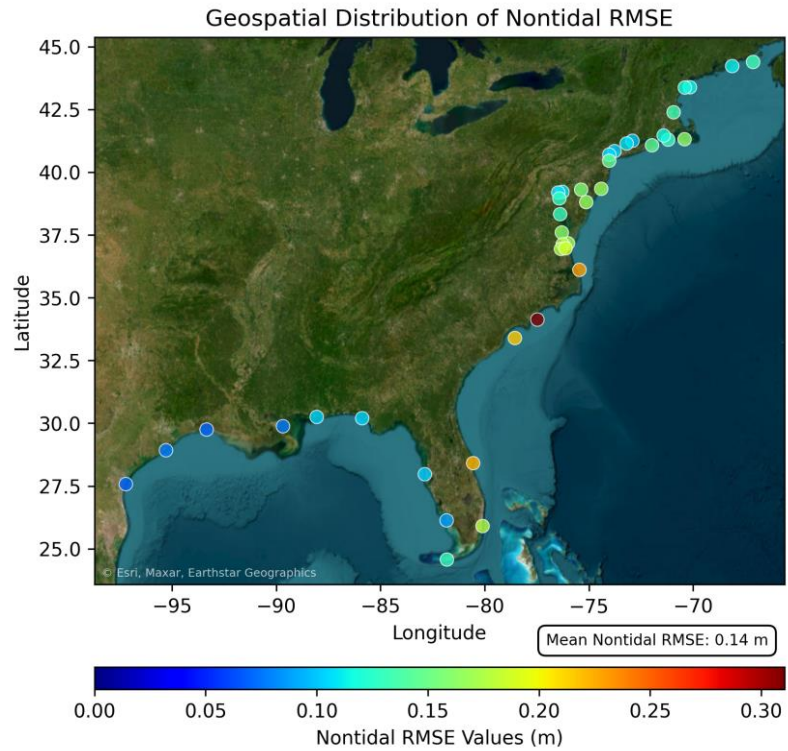

Figure 4. Spatial distribution of non-tidal water level RMSE for December 2016.

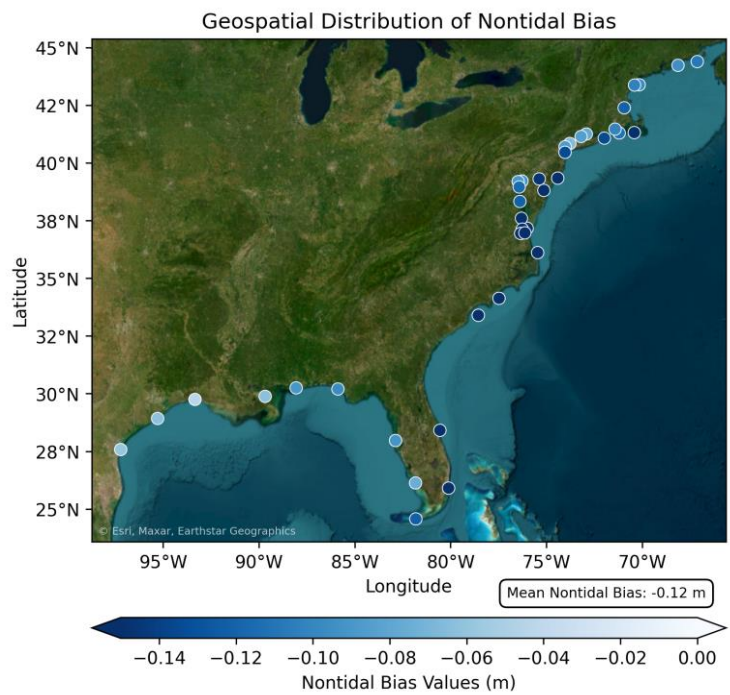

Figure 5. Spatial distribution of non-tidal water level bias for December 2016.

*Table 3. Calculated Bias and Root Mean Square Error (RMSE) of modeled Total Water Level (TWL) and Nontidal component for December 2016.*

| Station ID | TWL Bias (m) | TWL RMSE (m) | Nontidal Bias (m) | Nontidal RMSE (m) |
|---|---|---|---|---|
| 8411060 | -0.11 | 0.40 | -0.11 | 0.13 |
| 8413320 | -0.10 | 0.32 | -0.09 | 0.11 |
| 8418150 | -0.10 | 0.32 | -0.10 | 0.11 |
| 8423898 | -0.10 | 0.39 | -0.10 | 0.12 |
| 8443970 | -0.12 | 0.36 | -0.12 | 0.13 |
| 8447930 | -0.15 | 0.18 | -0.15 | 0.16 |
| 8452660 | -0.11 | 0.17 | -0.11 | 0.13 |
| 8454049 | -0.10 | 0.16 | -0.10 | 0.12 |
| 8465705 | -0.08 | 0.14 | -0.08 | 0.10 |
| 8467150 | -0.09 | 0.15 | -0.08 | 0.11 |
| 8510560 | -0.13 | 0.16 | -0.13 | 0.14 |
| 8516945 | -0.07 | 0.16 | -0.07 | 0.10 |
| 8518750 | -0.09 | 0.16 | -0.09 | 0.10 |
| 8531680 | -0.14 | 0.17 | -0.13 | 0.15 |
| 8534720 | -0.15 | 0.18 | -0.15 | 0.17 |
| 8537121 | -0.15 | 0.20 | -0.15 | 0.16 |
| 8557380 | -0.16 | 0.18 | -0.15 | 0.17 |
| 8573364 | -0.09 | 0.11 | -0.09 | 0.11 |
| 8574680 | -0.09 | 0.12 | -0.09 | 0.11 |
| 8575512 | -0.12 | 0.13 | -0.12 | 0.13 |
| 8577330 | -0.12 | 0.13 | -0.12 | 0.13 |
| 8632200 | -0.16 | 0.18 | -0.16 | 0.17 |
| 8636580 | -0.15 | 0.16 | -0.15 | 0.16 |
| 8637689 | -0.17 | 0.19 | -0.17 | 0.18 |
| 8638610 | -0.17 | 0.18 | -0.17 | 0.17 |
| 8638863 | -0.18 | 0.19 | -0.18 | 0.19 |
| 8651370 | -0.22 | 0.24 | -0.22 | 0.23 |
| 8658163 | -0.29 | 0.32 | -0.29 | 0.31 |
| 8721604 | -0.21 | 0.23 | -0.21 | 0.22 |
| 8722670 | -0.17 | 0.19 | -0.17 | 0.18 |
| 8724580 | -0.12 | 0.18 | -0.12 | 0.13 |
| 8726724 | -0.09 | 0.13 | -0.09 | 0.10 |
| 8725110 | -0.08 | 0.10 | -0.07 | 0.09 |
| 8729210 | -0.10 | 0.11 | -0.10 | 0.10 |
| 8735180 | -0.09 | 0.12 | -0.09 | 0.10 |
| 8761305 | -0.06 | 0.09 | -0.06 | 0.08 |
| 8768094 | -0.04 | 0.13 | -0.04 | 0.07 |
| 8772447 | -0.06 | 0.09 | -0.06 | 0.08 |
| 8775870 | -0.06 | 0.08 | -0.06 | 0.07 |

### 3.3 Total Water Level

The models' performance in terms of tidal and non-tidal components is a positive indicator regarding the total water level accuracy. Figures 6 and 7 illustrate, respectively, the spatial distribution of the total water level RMSE and bias for December 2016. The mean RMSE of modeled total water level is approximately 0.19 m, with the stations in the Gulf of Maine exhibiting total water level RMSE larger than 0.30 m. The minimum RMSE is approximately 0.08 m, and the maximum is 0.40 m (Table 3). The model is generally negatively biased during the simulation period, with a mean bias of approximately -0.12 m. The minimum bias is as low as -0.04 m while the maximum bias is about -0.29 m (Table 3).

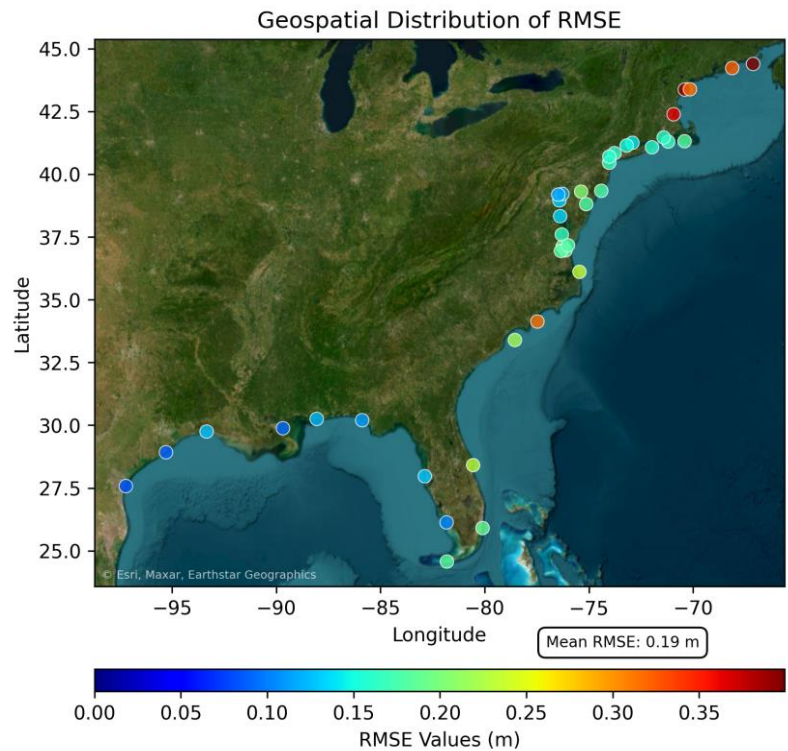

*Figure 6. Spatial distribution of total water level RMSE for December 2016.*

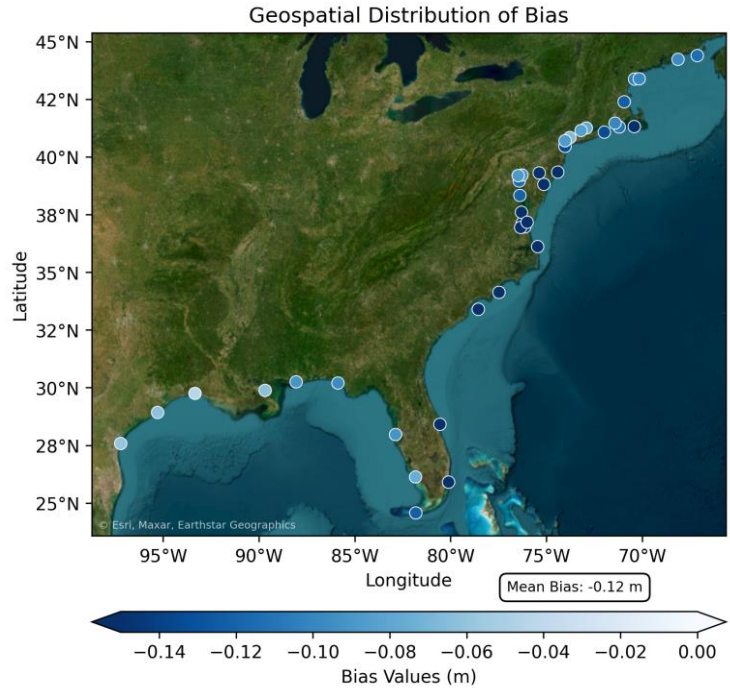

*Figure 7. Spatial distribution of total water level bias for December 2016.*

Overall, our results for tidal constituents and total water level demonstrated strong agreement with established research in continental-scale hydrodynamic modeling. The comprehensive tidal harmonic analysis shows that the model successfully captures both the amplitude and phase of dominant tidal constituents across the Western North Atlantic domain. For the principal lunar semi-diurnal M2 constituent, which dominates the East Coast stations, the model achieves excellent agreement with observations. Similarly, for the diurnal K1 constituent, prevalent in the Gulf of Mexico, the validation metrics indicate high model accuracy. Despite our considerably shorter simulation period, our simulated water levels from the coupled SCHISM and WWM III model compare favorably with the recent work by Cui et al., (2024), who utilized the stand-alone SCHISM model for predicting water levels along the U.S. East and Gulf Coasts. The model's ability to capture both deterministic tidal signals and meteorological influences is particularly encouraging for operational applications, including storm surge prediction and coastal flooding assessment.

### 3.4 Wave Characteristics

This section compares WWM-III results against buoy observations. We conduct the comparisons for three different simulations, activating a distinct wave physics parameterization in each one, namely "Ardhuin", "Makin and Stam", and "Cycle Three". Performance metrics focus on the accuracy of modeled wave height, period, and direction.

Figure 8 presents time series comparisons of significant wave heights at NOAA NDBC buoy stations over a selected simulation period. These comparisons reveal distinct performance patterns across the three wave physics parameterizations tested in WWM-III. Visual inspection of the time series shows that all three parameterizations capture the general temporal evolution of significant wave heights, including major peak events. However, differences in accuracy are evident across stations and wave conditions. All parameterizations consistently underestimate observed wave heights, which are denoted with black solid lines. More specifically, in terms of RMSE and bias, "Ardhuin" (blue dashed line) performed better at stations 41048, 42019, 42002, 42001, and 42003. Based on the same metrics, "Makin and Stam" (orange dashed line) performed better at stations 44027, 44017, and 44025. At buoy locations 41002 and 41049, we notice that both "Ardhuin" and "Makin and Stam" produced similar results in terms of RMSE; however, "Ardhuin" shows greater accuracy during peak wave events.

Our results reveal a distinct geographic pattern in parameterization performance. "Ardhuin" formulation generally demonstrates superior accuracy in capturing wave height variations across stations in the Gulf of Mexico, while the "Makin and Stam" parameterization performs more accurately along the U.S. East Coast stations. Several notable exceptions exist, particularly at stations 44005 and 44066, where the "Cycle Three" formulation unexpectedly outperforms the other parameterizations despite its earlier development. Another notable observation is that for offshore buoys located in deep waters beyond the continental shelf (such as stations 41048, and 41049), the "Ardhuin" parameterization demonstrates superior skill in capturing extreme wave height peaks, suggesting its greater suitability for modeling open ocean wave dynamics where wave-current interactions are less influenced by bathymetric complexities.

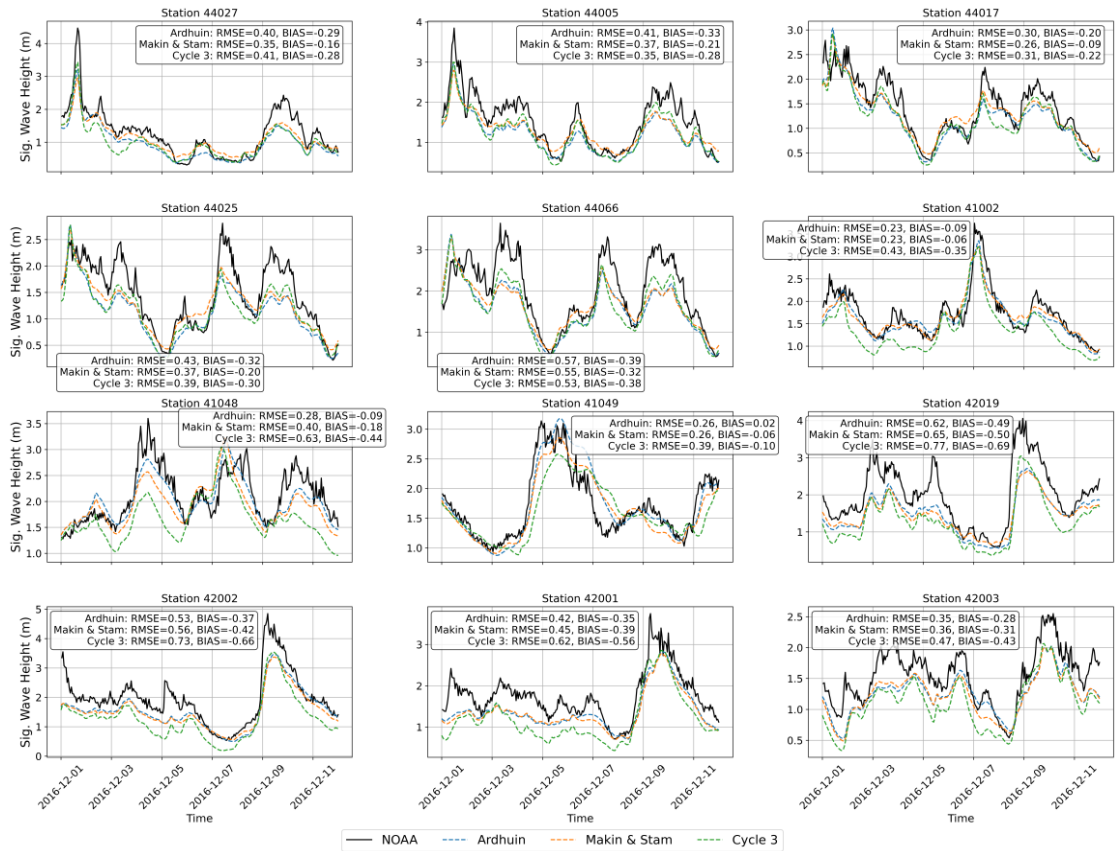

*Figure 8. Comparison of observed and modeled significant wave heights (in meters) at NOAA NDBC buoy stations during the first 12 days of December 2016. Black solid lines represent NOAA observational data, while colored lines show model outputs using three different wave physics parameterizations: "Ardhuin" (blue dotted), "Makin & Stam" (orange dashed), and "Cycle 3" (green dashed). Station identification numbers are shown at the top of each subplot. RMSE and bias for each parameterization with respect to the observed wave height are displayed within each panel.*

Figure 9 presents time series comparisons of mean wave periods at the same buoy stations and time period. These comparisons reveal significant differences in how the three wave physics parameterizations results in wave period characteristics. We notice that "Cycle Three" consistently greatly underestimates wave periods across all stations. "Makin and Stam" performs better in terms of RMSE and bias in all stations; however, "Ardhuin" is occasionally more accurate in capturing peak periods during certain events. In general, all models show limitations in capturing rapid transitions in wave period. Future investigation is needed to identify the underlying causes of these discrepancies and to potentially improve the existing wave physics parameterizations.

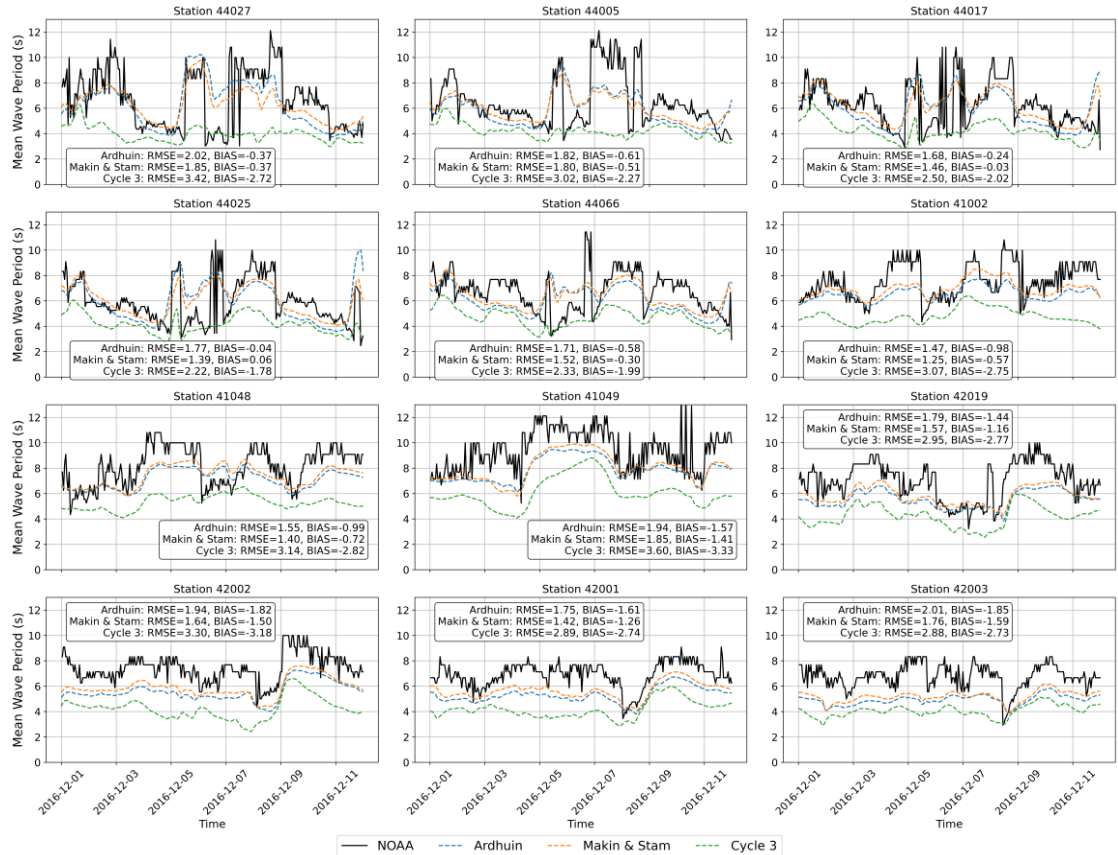

*Figure 9. Comparison of observed and modeled wave period (in seconds) at NOAA NDBC buoy stations during the first 12 days of December 2016. The notation followed is the same as Figure 8.*

Figure 10 presents comparisons of mean wave direction (in degrees) from which waves are coming from, measured clockwise from true north. Visual examination shows that all three parameterizations capture the broad directional shifts, but with varying degrees of accuracy. The nature of directional errors differs from those seen in the wave height and period. Rather than systematic under/over-estimation, directional errors often manifest as phase shifts or intermittent directional jumps. Unlike with wave height and period, "Cycle Three" performs remarkably well in terms of RMSE at stations 44027, 44005 44017, 44025, 44066, 41002, 41048, and 42019 while "Makin and Stam" displays a smaller RMSE at station 42002 and "Ardhuin" at stations 41049, 42001 and 42003.

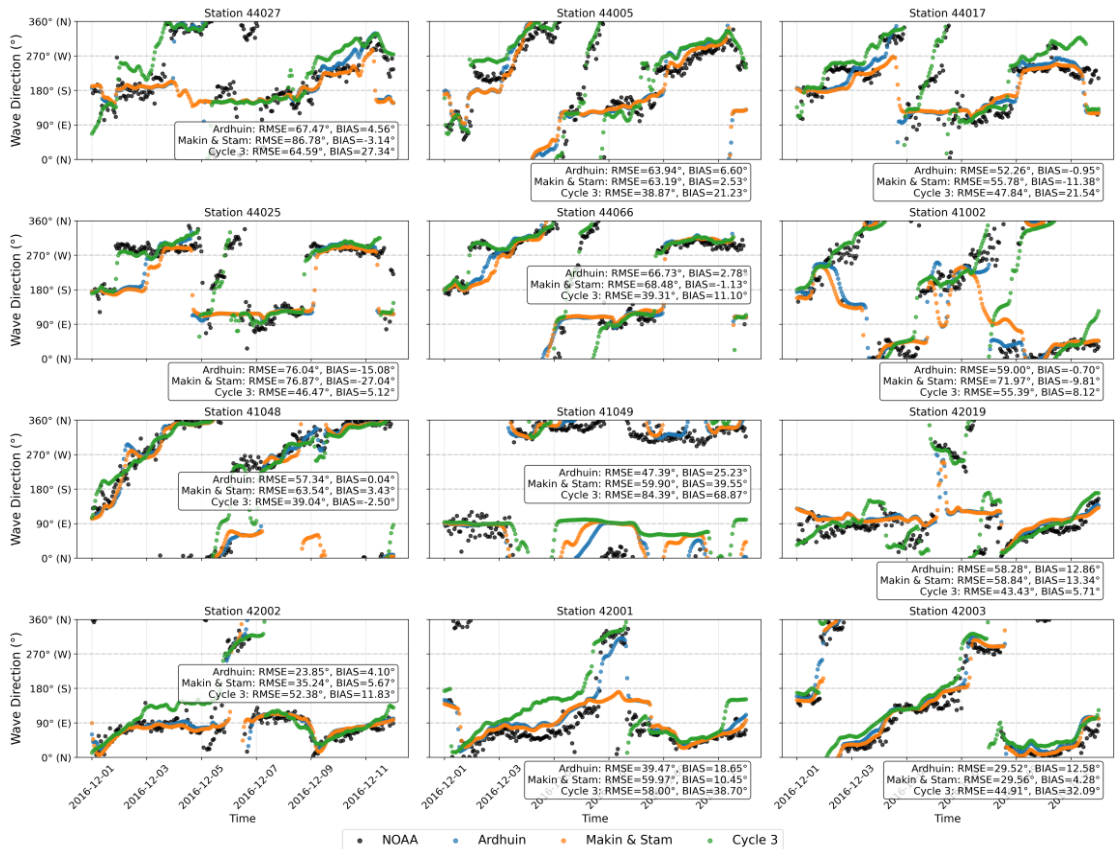

*Figure 10. Comparison of observed and modeled wave direction at NOAA NDBC buoy stations during the first 12 days of December 2016. The notation followed is the same as Figure 8. Wave direction is where the "waves are coming from".*

Overall, our evaluation of different wave physics parameterizations revealed distinct performance patterns that provide valuable insights for operational wave forecasting applications. The "Makin and Stam" parameterization showed greater accuracy along the U.S. East Coast for both significant wave heights and periods. In contrast, the "Ardhuin" formulation demonstrated improved performance, particularly in capturing extreme wave height peaks in the Gulf of Mexico and for deep ocean stations. Interestingly, the "Cycle Three" formulation, while consistently underestimating wave periods, performed well in predicting wave directions across most stations. These findings suggest that no single parameterization is optimal for all wave characteristics across different geographic and bathymetric settings, highlighting the importance of selecting appropriate physics configurations based on specific forecasting objectives and regional considerations.

## 4. Summary and Conclusion

In this study, we have implemented and evaluated a coupled three-dimensional baroclinic hydrodynamic and spectral wind-wave model, SCHISM+WWM III, for the Western North Atlantic Ocean. Our comprehensive validation against in-situ measurements from NOAA tide gauges and NDBC buoys demonstrated the model's robust capabilities in simulating water levels, tides, and wave characteristics throughout the domain. The model showed excellent skill in reproducing tidal constituents. This accuracy extends to non-tidal components and total water level predictions, confirming the model's ability to capture both deterministic tidal processes and meteorologically driven water level variations. Our systematic evaluation of different wave physics parameterizations revealed interesting performance patterns, which highlight the importance of selecting appropriate physics configurations based on specific forecasting objectives and regional characteristics. By incorporating full baroclinic dynamics and wave-current interactions through the coupled SCHISM+WWM III framework, the established modeling system is capable of resolving the complex interactions between tides, waves, atmospheric forcing, and oceanic circulation patterns that collectively drive water level and wave variabilities across multiple spatial and temporal scales.

This work establishes a validated modeling baseline that can be extended to explore various research questions, such as the influence of currents on wave spectra evolution. Through its improved accuracy and process representation, the model presented here offers potential applications in coastal management, maritime operations, and climate change adaptation planning throughout the Western North Atlantic region.

The computational mesh used in this study was adapted from STOFS-3D-Atlantic, a NOAA operational system that employs the stand-alone SCHISM model for coastal water level and flood forecasting. Our results show that coupling this model with a spectral wave model, WWM III in the present study, enables accurate simulation of both water levels and wave characteristics. Consequently, this modeling system has strong potential for operational wave forecasting within the STOFS-3D framework. Future work may further explore the system's performance when integrated with alternative spectral wave models, such as the widely used WAVEWATCH III and SWAN, to assess model sensitivity and optimize forecasting capabilities.

**Declaration of generative AI and AI-assisted technologies in the writing process**

During the preparation of this work the author(s) used Claude.ai and ChatGPT in order to correct grammatical or syntax errors, rephrase sentences to improve cohesion and coherence, and enhance clarity of the manuscript. After using this tool/service, the author(s) reviewed and edited the content as needed and take(s) full responsibility for the content of the publication.

**REFENCES**